\documentclass[preprint]{aastex}
\usepackage{natbib}
\shorttitle{An H-R-like diagram for galaxies}
\shortauthors{Feoli \& Mancini}
\received{2009 April 9}
\accepted{2009 August 6}

\begin{document}

\title{A HR-like diagram for galaxies: the $M_{\bullet}$ versus $M_{\mathrm{G}} \sigma^2$ relation}

\author{A. Feoli}
\affil{Dipartimento di Ingegneria, Universit\`{a} del
            Sannio, \\Corso Garibaldi n. 107, Palazzo Bosco Lucarelli  \\ 82100 -- Benevento, Italy.}
\email{feoli@unisannio.it} \and
\author{L. Mancini}
\affil{Dipartimento di Fisica "E. R. Caianiello",
Universit\`{a} di
            Salerno, \\via S. Allende  \\ 84081 -- Baronissi (SA), Italy.}
\email{lmancini@physics.unisa.it}

\begin{abstract}
We show that the relation between the mass of supermassive black holes located in the center of the host galaxies and the kinetic energy of  random motions of the corresponding bulges is a useful tool to study the evolution of galaxies. In the form $\log_{10}(M_{\bullet})=b+m \log_{10}(M_{\mathrm{G}}\sigma^2/c^2)$, the best-fitting results for a sample of 64 galaxies of various morphological types are the slope $m=0.80\pm0.03$ and the normalization $b=4.53\pm0.13$. We note that, in analogy with the Hertzsprung–-Russell diagram for stars, each morphological type of galaxy generally occupies a different area in the $M_{\bullet}-(M_{\mathrm{G}} \sigma^2)/c^2$ plane. In particular, we find elliptical galaxies in the upper part of the line of best fit, the lenticular galaxies in the middle part, and the late-type galaxies in the lower part, the mass of the central black hole giving an estimate of the age, whereas the kinetic energy of the stellar bulges is directly connected with the temperature of each galactic system. Finally, the values of the linear correlation coefficient, the intrinsic scatter, and the $\chi^2$ obtained by using the $M_{\bullet} - M_{\mathrm{G}} \sigma^2$ relation are better than the corresponding ones obtained from the $M_{\bullet}-\sigma$ or the $M_{\bullet}-M_{\mathrm{G}}$ relation.
\end{abstract}

\keywords{black hole physics -- galaxies: general -- galaxies: kinematics and dynamics --  galaxies: statistics}

\section{Introduction}
Today the fact that many galaxies, of different morphological types, host a supermassive black hole (SMBH) at their center has been established on quite solid grounds. The studies of the kinematics of galaxies and the combination of multi-band observations have played a major role in this scientific process. At the same time, the idea that the mass of a central SMBH is correlated with the evolutionary state of its host galaxy is being
consolidated among the scientific community. In order to qualify this correlation, many relationships have been proposed between the mass of the SMBHs and almost all the possible parameters of the host galaxy bulges: the velocity dispersion~\citep{ferrarese00,gebhardt00,tremaine02}, the bulge luminosity or
mass~\citep{kormendy95,marel99,richstone98,magorrian98,marconi01,merritt01,laor01,wandel02,gebhardt03,marconi03,haring04,gultekin09b}, the galaxy light concentration~\citep{graham01}, the X-ray power density spectra~\citep{czerny01}, the dark matter halo~\citep{ferrarese02}, the radio core length~\citep{cao02}, the effective radius~\citep{marconi03}, the Sersic index~\citep{graham05,graham07}, the inner core radius~\citep{lauer07}, the gravitational binding energy and gravitational potential~\citep{aller07}, the metal abundance~\citep{kisaka08}, the core mass deficit~\citep{kormendy09}, combination of bulge velocity dispersion, effective radius and/or intensity~\citep{aller07}, and, very recently, the X-ray luminosity, the radioluminosity~\citep{gultekin09c}, and active galactic nucleus jets~\citep{sokker09}.  An alternative approach has been proposed by \citet{feoli05}, who suggested a relation between the black hole (BH) mass and the kinetic energy of elliptical galaxies.
Then, these authors extended the study of the relation  $M_{\bullet}$ versus $M_{\mathrm{G}} \sigma^2$ also to lenticular and spiral galaxies~\citep{feoli07}, enlarging their sample to a total of 29 galaxies, and finding that it has a scatter smaller than the most famous $M_{\bullet} - \sigma$ relation. Here, $M_{\mathrm{G}}$ is the bulge mass of the galaxies, where \emph{bulge} as usual refers to either the hot, spheroidal component (no dark halo or disk contribution) of a spiral/lenticular galaxy or to a full elliptical galaxy~\citep{aller07,cattaneo09}. In particular, for late-type galaxies, $M_{\mathrm{G}}$ is the mass enclosed within a sphere of radius $R$ fixed from the surface-brightness profile. As an example, \citet{aller07} consider that $R=10 \,R_{\mathrm{e}}$ is ``the best substitute for the bulge mass'', with $R_{\mathrm{e}}$ being the effective radius of a galaxy.

Actually, the main problem is that almost all the above-quoted relations are very tight, so it is very difficult to find, by studying the scatter of each one of them, the ``most fundamental one''~\citep{tremaine02,novak06,gultekin09b}. Without definitely solving this hard problem, the attention of an increasing number of scientists is now focused on the $M_{\bullet} - \sigma$ law in order to study the behavior of some peculiar subsets of galaxies. This led to discover that the line of the best fit of that relationship is different for barred galaxies with respect to the barless ones~\citep{graham08}. The same occurs for bulges and pseudo-bulges~\citep{hu08,gadotti09}, core or coreless~\citep{hu08}, active or quiescent~\citep{barth05,wyithe06a,wyithe06b,zhang08,greene06}.

At the same time, from the theoretical point of view, a lot of interesting (analytical and semianalytical) models were constructed to explain the experimental results (see, for example,~\citet{haehnelt00,burkert01,wyithe02,wyithe03,dokuchaev03,volonteri03,miller06,croton06,delucia07}). It is well known that a useful method to obtain theoretical predictions, which can be compared with the correlations derived from experimental data, is based on numerical simulations. \citet{hopkins07} examined the origin and the evolution of the correlations between the properties of SMBHs and their host galaxies using hydrodynamical simulations of major galaxy mergers, including the effects of gas dissipation, cooling, star formation, and BH accretion and feedback. Their simulations suggest the existence of a SMBH \emph{fundamental plane}, analogous to the fundamental plane of elliptical galaxies. The best relation that they found (the one with the lowest scatter) is
\begin{equation}
\log_{10}(M_{\bullet}) = (7.93 \pm 0.06) + (0.72 \pm 0.12) \log_{10}(M_{11}^*) +
(1.40 \pm 0.49) \log_{10}( \sigma_{200}),
\label{Eq_01}
\end{equation}
where $M_{11}^*$ is the galaxy stellar mass in units $10^{11}$ $M_{\sun}$, and $\sigma_{200}$ is the bulge velocity dispersion in units of $200$ km sec$^{-1}$. These authors also show the main role played by the kinetic energy of random motions first proposed by \citet{feoli05}. In particular, they say: ``\emph{we therefore naively expect that the BH mass should scale with $M_* \sigma^2$}'', and declare that the correlation between the BH mass and the $M_* \sigma^2$ ``\emph{is in some sense more basic then the correlation between the BH mass and $M_*$ or $\sigma$}''. In other words, their fundamental plane in BH mass can be well represented as a ``tilted'' correlation between BH mass and the kinetic energy of the random motions in the host galaxies (see Figure 10 of \citet{hopkins07}). Another clue is the ratio between the coefficients in Equation~(\ref{Eq_01}) multiplying the $\log_{10}(M_{11}^*)$ and $\log_{10}(\sigma_{200})$ which is very close to 0.5. This is also remarked by \citet{marulli08}, who modeled the cosmological co-evolution of galaxies and their central SMBHs within a semianalytical framework. Their model matches well enough the SMBH fundamental plane relation derived by \citet{hopkins07}, and their conclusion is identical: the SMBH mass does not simply scale with the star formation (stellar mass) or the velocity dispersion of the host galaxy.

The results of \citet{hopkins07} and \citet{marulli08} give a strong evidence that galaxy spheroids and SMBHs do not form and evolve independently and support the approach of \citet{feoli05,feoli07}, who pointed to the relationship between the masses of the SMBHs and the kinetic energy of random motions in their host galaxies. The consequences for the theoretical models of SMBH growth and evolution are non-trivial.

In the present paper we want to extend the previous analysis of \citet{feoli05,feoli07} (hereafter, Paper I and Paper II) to a new set of 64 galaxies, almost all extracted by the catalogue of \citet{graham08}. The main aim is to probe if the $M_{\bullet} - M_{\mathrm{G}} \sigma^2$ relation is really a helpful instrument to study the evolution of the galaxies, that is if it can play the same role as the Hertzsprung–-Russell (H–-R) diagram in the description of the evolution phases of stars. We will see that different morphological types of galaxies occupy different positions in the $M_{\bullet} - M_{\mathrm{G}} \sigma^2$ plane, reflecting their age and intrinsic features. Finally, we want to confirm that the linear correlation coefficient, the intrinsic scatter, and the $\chi^2$ of our
relationship are better than the corresponding values for the $M_{\bullet}-\sigma$ or the $M_{\bullet} - M_{\mathrm{G}}$ law.  Our paper  is structured as follows. In \S~2 we define the samples used in our statistics. In \S~3 we explain our results and, finally, in \S~4 we draw our conclusions.
%

\section{The samples}
In order to have a homogeneous set of data, we have considered as the main reference for the masses of SMBHs and the velocity dispersions of the galaxies the catalogue published by \citet{graham08}. Alternative values (almost compatible within uncertainties) can be found for instance in \citet{hu08} or in the more recent paper of \citet{gultekin09b}. Our choice involves the values of the central velocity dispersions $\sigma_{\mathrm{c}}$, in contrast with our two previous papers where we used the effective dispersion velocity $\sigma_{\mathrm{e}}$. However, as already noted by \citet{novak06}, the two ways of measuring the velocity dispersion does not generate profound differences. This is also supported by the study of \citet{hu08}, who compared the effective dispersion $\sigma_{\mathrm{e}}$ with the central one $\sigma_{\mathrm{c}}$, finding that the differences are much smaller than their measurement errors. Also \citet{gultekin09b} compared $\sigma_{\mathrm{e}}$ to $\sigma_{\mathrm{c}}$ finding no systematic bias to high or low values. We also remark the fact that in our previous papers the data were extracted only by single sources: in Paper I all the values of the galaxies masses have been taken from \citet{curir93}, the velocity dispersions from \citet{busarello92}, and the SMBH masses from \citet{tremaine02}; in Paper II the three sets of data have been taken from \citet{haring04}. Here, due to the enlargement of the sample, a homogeneous choice is no more possible and we are forced to build up a sample of data from various catalogues -- essentially those of \citet{graham08},  \citet{haring04} and \citet{cappellari06} -- and single papers. A clear limit of a collection of data of this kind is, of course, related to the many different techniques utilized to estimate the masses of the bulges (dynamical or virial masses, Schwarzschild models, Jeans equation, etc.; see Appendix~\ref{appendixB} for a more comprehensive discussion), and of the SMBHs (gas or stellar kinematics, water maser, proper motions, etc.).

In this paper, we consider two samples of galaxies. The first sample (sample A) is composed by 49 galaxies included in the table 1 of \citet{graham08}. Actually, his catalogue is formed by 50 galaxies that are considered to have reasonable measurements of their SMBH masses. We exclude the galaxy IC2560, since a reliable value for its bulge mass is not available~\citep{ishihara01,schulz03}. In several cases we would have liked to substitute some data of the Graham catalogue with other measures which are less uncertain or simply more recent (like that of \citealt{gultekin09b}), but we did not do so in order to avoid the risk that the tightness of our relation might depend on a suitable choice of the data. For example, we have used for the mass of SMBH in the Milky Way the value cited by Graham even if we know that an update value is now available~\citep{gillessen09}, and we have included in the sample also the elliptical NGC221, which we would have liked to exclude from the fit as already done in Papers I and II. Starting from the Graham's catalogue we have fixed the total number and the names of galaxies, their velocity dispersions, the SMBH masses, and morphological types. In this way, only the galaxy mass remains as a free parameter but our choice was anyway restricted by using the data published by \citet{haring04} and \citet{cappellari06}.

The second sample (sample B) is composed by the galaxies of sample A plus other 15 galaxies whose parameters have been taken from table 2 of \citet{graham08} and from other papers. Of course, this enlarged sample does not have the aim to include all the galaxies with a measured BH mass or with an upper limit on its value. Sixty-four galaxies are listed in Table~\ref{Tab_01} and compose sample B, whereas only the first 49 galaxies are included in sample A.

Concerning the errors in the measures, we adopt the same strategy as in Paper II. Following \citet{haring04}, we consider that the error for the bulge mass is 0.18 dex in $\log_{10}M_{\mathrm{G}}$ for all the galaxies, while the relative error on the velocity dispersions is $10\%$.
%

\section{Results}
\begin{figure}
\centering{\includegraphics{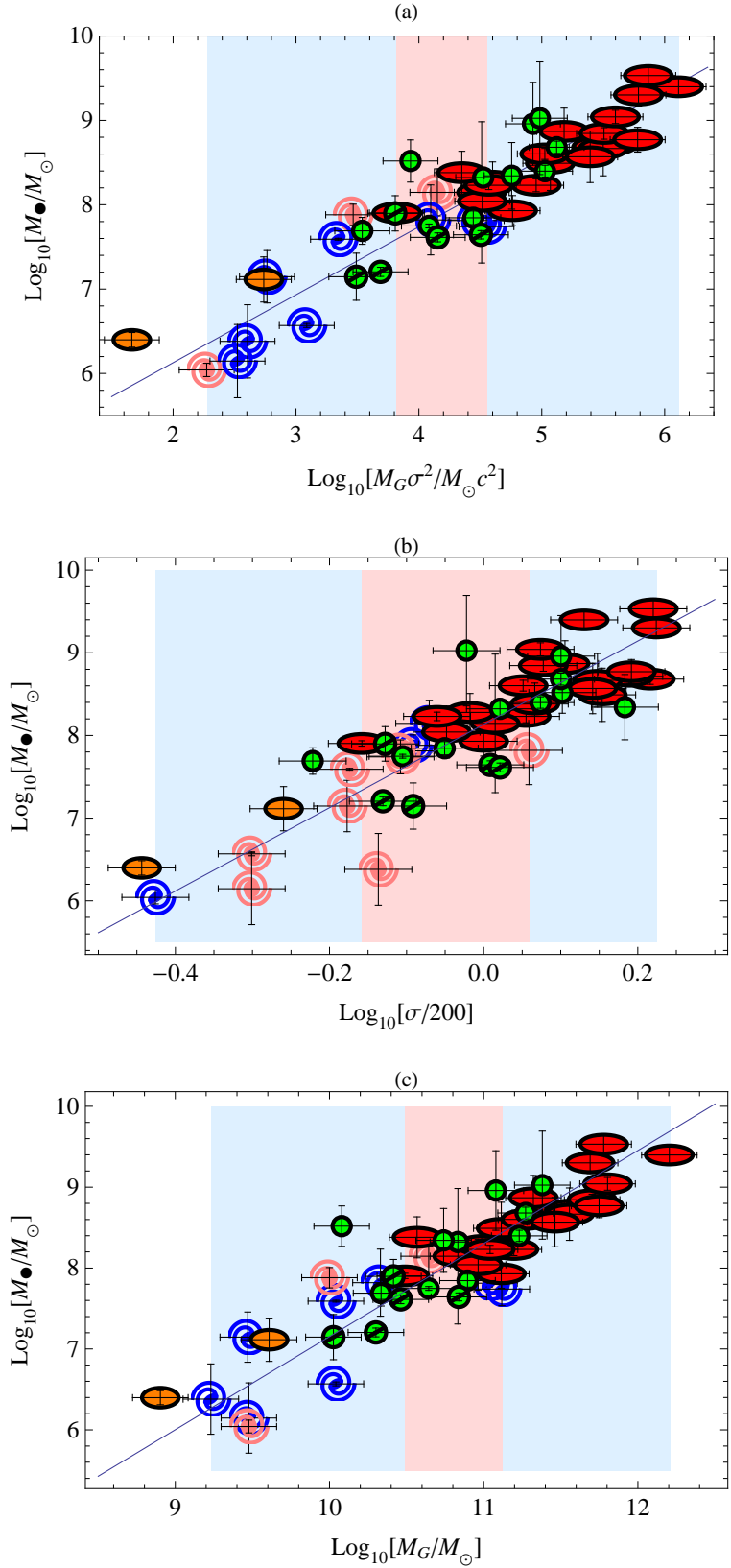}}
   \caption{Best-fitting (a) $M_{\bullet} - M_{\mathrm{G}} \sigma^2$, (b) $M_{\bullet} - \sigma$, and (c) $M_{\bullet} - M_{\mathrm{G}}$  relations for the elliptical galaxies (red ellipses), lenticular galaxies (green circles), barred lenticular galaxies (green barred circles), spiral galaxy (pink spirals), barred spiral galaxies (blue barred spirals), and dwarf elliptical galaxies (orange round ellipses) of sample A. The light-red colored area represents the transition area, in which the elliptical and the spiral galaxies are mixed together (see the text).}
 \label{Fig_01}
\end{figure}

The  relation between the mass of the SMBHs and  the kinetic
energy of random motions of the corresponding host galaxies has
been presented in Papers I and II in the form
\begin{equation}
\log_{10}(M_{\bullet}) = b + m \log_{10}(M_{\mathrm{G}}\sigma^2/c^2).
\label{Eq_02}
\end{equation}
Thus,  this relation can be used to predict the values of $M_\bullet$ in other galaxies
once we know their mass and velocity dispersion. In order to minimize the scatter in the
quantity to be predicted, we have to perform an ordinary least-squares regression of
$M_\bullet$ on $M_{\mathrm{G}}\sigma^2$ for the galaxies in Table~\ref{Tab_01}, of which
we already know both the quantities. In Table~\ref{Tab_02}, we compare the fits of our
relationship for the two samples and the corresponding fits for $M_{\bullet}-\sigma$ and
$M_{\bullet}-M_{\mathrm{G}}$ laws. As in Paper II (see also \citet{graham07}), these fits
were obtained taking into account the error bars in both variables and using the routine
FITEXY~\citep{press92} for a relation $y=b+mx$, by minimizing the $\chi^2$ (see
Appendix~\ref{appendix}). Comparing the results of the three laws, we notice that the
$\chi^2$, the intrinsic dispersion $\varepsilon_0$ (i.e., dispersion due to the galaxies
themselves rather than to measurement errors), and the Pearson linear correlation
coefficient $r$ of our relationship are better than the other ones (Table~\ref{Tab_02}).
This is also evident by the comparison of Figure~\ref{Fig_01}a, \ref{Fig_01}b and
\ref{Fig_01}c, where the three relations are reported in log-log plots (we associated a
particular marker to each galaxy according with its morphological type).  It is visually
clear that the galaxies (especially the spirals) are more spread in the case of the
$M_{\bullet}-\sigma$ or the $M_{\bullet}-M_{\mathrm{G}}$ relations, than in the
$M_{\bullet}-M_{\mathrm{G}}\sigma^2$ law. This result is also supported by the analysis
of \citet{gadotti09} who found that elliptical, classical bulges and pseudo-bulges follow
different relations between their stellar masses and velocity dispersions.

Comparing the results in Table~\ref{Tab_02} with the corresponding ones in table 3 of
Paper II, we observe that, by enlarging the sample, the correlation coefficient of our
relationship increases, showing the robustness of our idea. Actually, if the existence of
the correlation is not a novelty, since it has already been  found by \citet{feoli05},
the increase of the correlation coefficients with the enlarging of the sample is a result that had not at all been taken for granted.

As already observed by~\citet{novak06}, the question  ``which relation is better than the
others?'' is extremely sensitive to inaccurate estimates of the measurement errors. So,
the result that our $\chi^2$ is better than the $M_{\bullet}-\sigma$ and the
$M_{\bullet}-M_{\mathrm{G}}$ laws can be caused by an overestimation of the error on the
galaxy masses. In order to avoid a similar misleading result, we have checked what
happens using a standard least squared fitting, assuming that errors in the kinetic
energy are zero and that errors in the $\log_{10} M_{\bullet}$ are the same
$\epsilon_{\mathrm{y}}$ for each galaxy. The results are reported in Table~\ref{Tab_03}
and show that the scatter of our relation is better than the $M_{\bullet}-\sigma$ and the
$M_{\bullet}-M_{\mathrm{G}}$ laws even in this extreme case. Furthermore, the slope of
the line of best fit $m = 0.73 \pm 0.04$ is the same, inside the errors, as the one of
\citet{hopkins07} in Equation~(\ref{Eq_01}).

We note also that the slope of the $ M_{\bullet}-\sigma$ law depends on the errors and on
the fitting methods used, more than the other relations do. While the values in
Table~\ref{Tab_03} are close to the estimates of \citet{tremaine02} and
\citet{gultekin09b} (even if they have been obtained with a different sample), the values
in Table~\ref{Tab_02} are closer to the ones obtained by \citet{graham08} (he found
$m=5.22\pm0.40$ and $b=8.13\pm0.06$) with the same sample but with the \citet{akritas96}
method.

\begin{figure}
\centering{\includegraphics{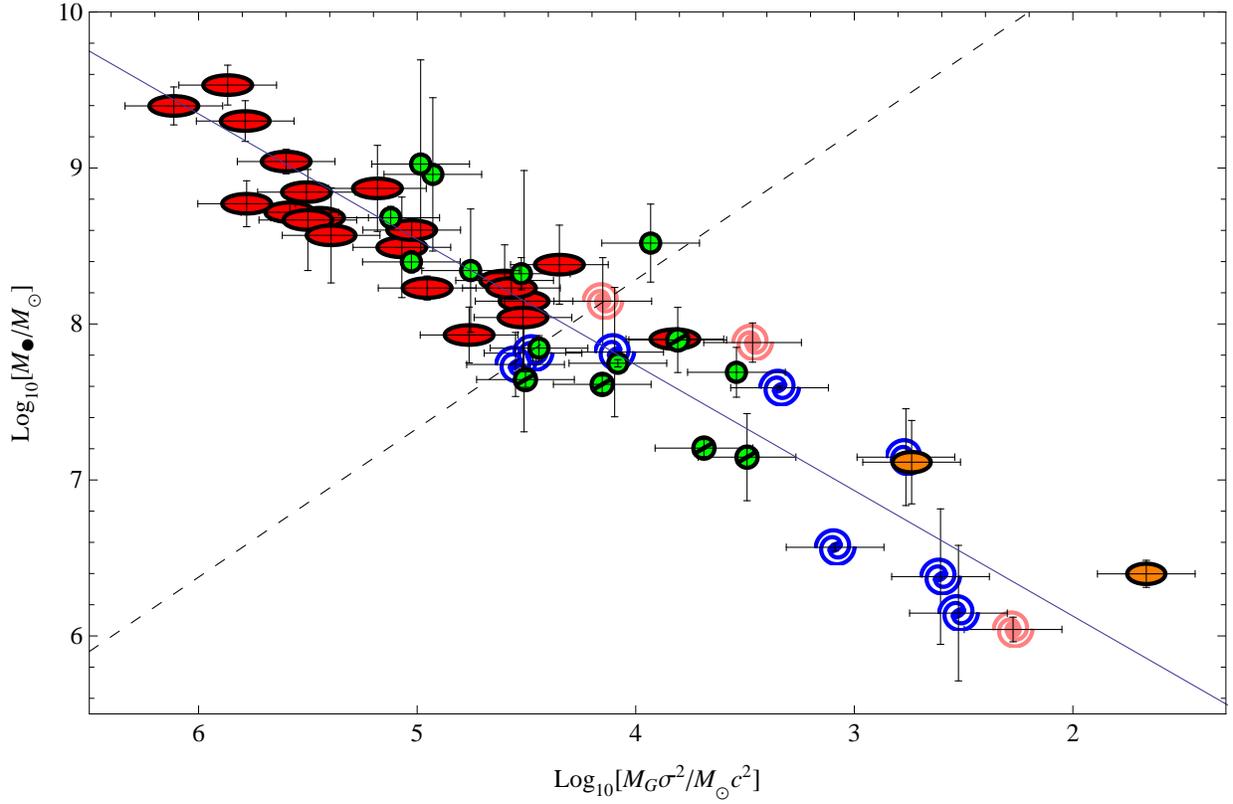}}
   \caption{For the $M_{\bullet} - M_{\mathrm{G}} \sigma^2$ relation, the line of best fit for the galaxies of sample A is represented by a solid line, while the dashed line separates the early type galaxies from the late type ones. The markers are the same of Figure~\ref{Fig_01}.}
 \label{Fig_02}
\end{figure}

A surprising result is shown  in Figure~\ref{Fig_02}, where we
performed a log-log plot of the energy stored by the SMBH,
$E_{\mathrm{st}}= M_{\bullet} c^2$, as a function of the bulge
kinetic energy of random motions, both normalized by the rest
energy of the Sun, $M_{\sun}c^2$. Given the line of best fit
(solid line) and a sort of border line (dashed line) that divides
the diagram in two parts, it is evident that:
\begin{enumerate}
\item almost all the elliptical galaxies (except NGC3377) are in the higher part of the diagram (over the dashed line),

\item the lenticular galaxies are located in the middle-upper part of the diagram,

\item the barred lenticular galaxies are located in the middle part of the
diagram (but under the dashed line),

\item all the spirals are in the middle-lower and in the lower parts of the diagram (under the dashed line),

\item in the lower part of the diagram we find also two dwarf elliptical galaxies: NGC221 and NGC4486A.

\end{enumerate}

In analogy with the H-R diagram for stars, each morphological type
of galaxy occupies a different area in the $M_{\bullet} -
(M_{\mathrm{G}} \sigma^2)/c^2$ plane. This effect is in part due
to the well known fact that $M_{\bullet}$ and $M_{\mathrm{G}}$
(also $\sigma$, even if with a lot of exceptions) generally
increase with the morphological type, but it is remarkable and not
granted that they simultaneously increase just in the right way to
produce the effect and a law with a minimal scatter. Compared with
the other two relations, this clear trend can be lightly
recognized also in Figure \ref{Fig_01}c, but it is not clear in
Figure \ref{Fig_01}b. For example, the ellipticals in Figure
\ref{Fig_01}a are more separated from the spirals with respect to
Figures \ref{Fig_01}b or \ref{Fig_01}c. This can be
quantified calculating the width of the transition area (light red
colored), in which the elliptical and the spiral galaxies are
mixed together, with respect to the entire area (light blue
colored) occupied by all galaxies (we exclude the dwarf
ellipticals). In Figure \ref{Fig_01}a, the red zone is only the
$19\%$ of blue area, whereas it is $33\%$ and $21\%$ in Figure
\ref{Fig_01}b and \ref{Fig_01}c, respectively.

\begin{figure}
\centering{\includegraphics{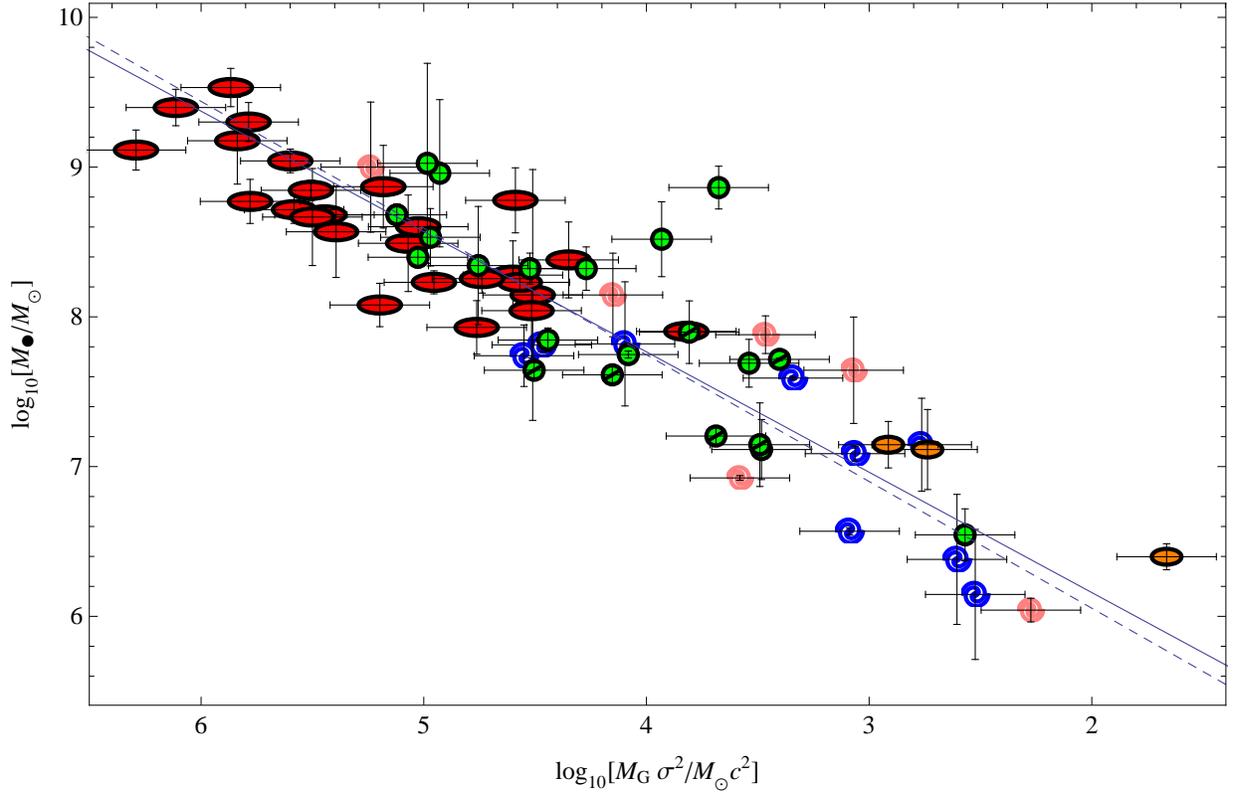}}
   \caption{For the $M_{\bullet} - M_{\mathrm{G}} \sigma^2$ relation, the line of best fit for the galaxies of sample B is represented by a solid line, whereas the one obtained without considering the three dwarf ellipticals by a dashed line. The markers are the same of Figure~\ref{Fig_01}.}
 \label{Fig_03}
\end{figure}

The general trend observed in Figure~\ref{Fig_02} is respected
in Figure~\ref{Fig_03} for the galaxies of sample B, even if
two lenticular galaxies appear in the lower part of the diagram
and a spiral galaxy in the upper part. The latter is the famous
\emph{Sombrero galaxy} (NGC4594), one of the largest galaxies in
the nearby Virgo Cluster, classified as a lenticular
by~\citet{magorrian98}. It is well known that it has a bright
nucleus and an unusually large classical bulge, testified by a
relatively large number of globular clusters. We know that the
classical bulges are believed to be generated by mergers and are
common in early type galaxies but become progressively rare
toward later types. They share some structural, dynamical, and
population properties with the lower-luminosity ellipticals
\citep{freeman07}. Actually, NGC4594 is surrounded by a halo of
stars, dust, and gas that indicate it may actually be described as
an elliptical galaxy that contains a more robust interior
configuration. Therefore, its presence in the upper part of the
diagram is not so improper. Later type galaxies like the Milky Way
mostly have small boxy bulges and are all in the lower part of the
diagram. On the other hand, both a classical bulge and an inner
boxy bulge are present in NGC224 (Andromeda galaxy, M31)
\citep{athanassoula06,beaton07}, which is located just in the
middle region.

Both in Figure~\ref{Fig_02} and \ref{Fig_03}, the
elliptical galaxies are all clustered very near the line of best
fit. Conversely, the galaxies of the other morphological types
look slightly more scattered. This is particularly true for the
lenticular galaxies in the middle-upper part of the diagram. Among
them, the galaxy NGC4342 is located quite far from the
best-fitting line. As already noted by \citet{cretton99}, NGC4342
is one of the galaxies with the highest SMBH mass to bulge mass
ratio. The consequent hypothesis that we are in the presence of a
galaxy in a particular evolutionary state is also supported by the
presence of both an outer disk and a stellar nuclear disk
\citep{bosch97}. Instead, the position of the lenticular galaxy
NGC7457 in the lower zone of the diagram is due to its SMBH which
is one of the least-massive BHs yet detected in the core
of a galaxy, roughly the same mass as the BH at the center
of our Galaxy. Equally, the boxy-bulge lenticular NGC7332 is
located in the middle-lower region of Figure~\ref{Fig_02}.
Going on, the peculiar galaxy NGC5128 (Centaurus A) appears in the
middle of the diagram, but still quite far from the other
lenticulars. The strange morphology of Centaurus A is generally
recognized as the result of a merger of two smaller galaxies. In
this way, it is possible to explain a bulge comprised mainly of
evolved red stars and a dusty disk, which has been the site of
recent star formation \citep{israel98}.

Finally, we note the presence of the intermediate-size elliptical
NGC3377 in the center of the graphic in Figure~\ref{Fig_02}, and
three dwarf elliptical galaxies in the lower part. Two of them are
NGC4742 and NGC4486A, both belonging to the Virgo Cluster of
Galaxies, whereas the small NGC221 (M32) is a satellite of M31.
Continuing the analogies with the H-R diagram, we can look at this
area as reserved to the dwarf ellipticals, in the same manner as
the region occupied by the white dwarfs in a classical
color-magnitude diagram. If the three dwarf ellipticals do not
really belong to the ``principal sequence'', we can exclude them
from the fit. In this case, the slope and the normalization in our
relationship for the reduced sample of 61 galaxies are
$0.84\pm0.03$ and $4.33\pm0.15$, respectively (the dashed line in
Figure~\ref{Fig_03}). Furthermore, comparing again the
$\chi_{\mathrm{r}}^2=1.80$ and the $r=0.91$ of our relation with
for instance the corresponding $\chi_{\mathrm{r}}^2=2.37$ and
$r=0.83$ for the $M_{\bullet}-\sigma$ law, we find an increase of
the gap between the two relations.

Also remarkable is the fact that all the barred galaxies
(lenticulars and spirals) are located only in the
lower/middle-lower part of the diagram (under the dashed line of
Figure~\ref{Fig_02}).

We also study a possible correlation between the activity
(Seyfert, Liner, etc.) of each galaxy with its position on the
diagram, but we did not note any particular trend.

%
\section{Discussion}
In this paper, we have investigated the relation between the mass
of  the SMBHs and the kinetic energy of the random motion of the
corresponding galaxies. This relation has been tested on a
homogeneous sample of 49 galaxies and then on a more enlarged
sample of 64 ones. As shown in Table~\ref{Tab_02}, the statistical
analysis confirms the result of our two previous papers, that is
the proposed relation works well and better than the most common
$M_{\bullet} - \sigma$ law or the $M_{\bullet} - M_{\mathrm{G}}$
one. Furthermore, the main result that we report consists in the
particular positions of the galaxies in the $M_{\bullet} -
(M_{\mathrm{G}}\sigma^2)/c^2$ plane, which resembles the H-R diagram
for the stars. Other analogies also exist between the two diagrams.
The H-R diagram connects the energy radiated (per unit
time) by the nucleus of a star with its surface temperature. In
the same way, our diagram connects a property of the inner nucleus
of a galaxy, the energy stored by the SMBH, $M_{\bullet}c^2$, with
a property of the external surface of its bulge, i.e., the kinetic
energy of random motions. This energy is related just with the
temperature of the stellar system. In fact, let us consider a
spherically symmetric distribution of stars with density $\rho$,
whose dynamical state is described by a distribution function of
the form
\begin{equation}
F(E)=\frac{\rho}{\left(2 \pi \sigma^2\right)^{3/2}}e^{E/\sigma^2},
\label{Eq_3}
\end{equation}
where $E=\Psi-v^{2}/2$ is the binding energy,  and $\Psi$ is the relative gravitational
potential~\citep{binney87}. Now, it is well known that the structure of a collisionless
system of stars, whose density in the phase space is given by Equation~(\ref{Eq_3}), is
identical to the structure of an isothermal self-gravitating sphere of gas, if we set
\begin{equation}
M \sigma^2=N k_{\mathrm{B}}T,
\end{equation}
with $M$ being the total mass of the system,  $N$ the number of objects
contained in the system, $k_{\mathrm{B}}$ the Boltzmann's
constant, and $T$ the temperature of the system. Since a stellar bulge can be considered, with a good
approximation, similar to a spherically symmetric system, its
kinetic energy of the random motions $M_{\mathrm{G}}\sigma^2$
gives an indication of the temperature of the galaxy bulge.

On the other hand, since the SMBH at the center of galaxies can only
increase its mass, the stored energy is directly connected with
the initial density of the system and its evolutionary state. So,
the stored energy of central SMBH will guide the galaxy along the
evolutionary process, and, in that sense, an accretion of the
SMBH, bound up with the flow of time, will imply a migration of
the galaxy position from the lower-right part of the $M_{\bullet}
- (M_{\mathrm{G}} \sigma^2)$ diagram to the upper left. This
migration does not involve a brutal transformation of a spiral
galaxy in an elliptical one but, since we consider just the mass
of the bulge in our relation, we suppose that the spheroidal
components of the spirals increase their size becoming similar to
ellipticals, as it happened for the Sombrero galaxy discussed in
the previous section.

\acknowledgements
We are grateful to Federico Marulli and Gianni Busarello  for useful discussions and to Michele Cappellari for a private communication about the values of some galaxy masses. We also thank the anonymous referee for its suggestions and comments that have
helped us to improve the quality of this paper. This research was partially supported by FAR fund of the University of Sannio. L.M. acknowledges support for this work by MIUR through PRIN 2006 Protocol 2006023491\_003, by research funds of the Italian Space Agency, by funds of Regione Campania, L.R. n.5/2002, year 2005 (run by Gaetano Scarpetta), and by research funds of the University of Salerno.

%
\appendix
\section{Appendix}\label{appendix}
The formula used in this paper to estimate the maximal errors
in the functions F of the parameters $(a,b,c,....)$ is
\begin{equation}
\Delta F(a,b,c) = \left| \frac{\partial F}{\partial a}\right|
\Delta a + \left|\frac{\partial F}{\partial b}\right| \Delta b +
\left|\frac{\partial F}{\partial c}\right| \Delta c.
\end{equation}
The reduced $\chi^2$, used in Table~\ref{Tab_02}, is defined as
\begin{equation}
\chi^2_r = \frac{\chi^2}{N-2} = \frac{1}{N-2} \sum_{i=1}^{N} \frac{(y_i
-b-m x_i)^2}{(\Delta y_i)^2 + m^2 (\Delta x_i)^2},
\end{equation}
for a relation of the form $y = b + m x$, where $N$ is the number
of galaxies in the sample. The internal  scatter $\varepsilon_{0}$
in Table~\ref{Tab_02} is calculated for fixed values of $m$ and
$b$ imposing that
\begin{equation}
\chi^2_r = \frac{1}{N-2} \sum_{i=1}^{N} \frac{(y_i
-b-m x_i)^2}{(\Delta y_i)^2 + \varepsilon_{0}^2+ m^2 (\Delta x_i)^2} = 1.
\end{equation}
Following \citet{tremaine02}, the internal
 scatter can be computed also   replacing the
  error $\Delta y$ with $(\Delta y^2 + \varepsilon_{0}^2)^{1/2}$ before the fitting procedure, and then
  adjusting $\varepsilon_{0}$ and refitting until the $\chi^2_r$ is equal to 1. This approach affects the values of the
  slope and the normalization of the best-fitting relations. We reported them in
  Table~\ref{Tab_04} together with the values of $\varepsilon_{0}$ for the galaxies of sample A.

Furthermore, the $\epsilon_{\mathrm{y}}$ used in Table \ref{Tab_03}, is defined as
\begin{equation}
\epsilon_{\mathrm{y}}^2 = \frac{1}{N-2} \sum_{i=1}^{N} (y_i
-b-m x_i)^2.
\end{equation}

Finally, the Pearson linear correlation coefficient is
\begin{equation}
r = \frac{\sum ^n _{i=1}(x_i - \bar{x}) - (y_i - \bar{y})}{\sqrt{\sum ^n _{i=1}(x_i - \bar{x})^2} \sqrt{\sum ^n _{i=1}(y_i - \bar{y})^2}} \;.
\end{equation}

\section{Appendix}\label{appendixB}
There are  several methods to estimate the mass of a galaxy. Some
of them are based on the virial theorem, other on the Jeans
equation or on self-consistent models. When we are interested in
the mass of a whole galaxy, then it is possible to use the virial
theorem in its scalar form $2T + U = 0$ and, knowing the rotation
velocity and velocity dispersion, the corresponding virial mass
can be derived~\citep{busarello90}. This approach often reduces to
calculate the dynamical mass
\begin{equation}
M_{\mathrm{dyn}}=\frac{k \sigma^2 R_{\mathrm{e}}}{G}.
\end{equation}

Instead, the Jeans equation is particularly useful to calculate
the mass of late type galaxies,  where we must take into account
only the spheroidal component. If one wants to study only the part
of a galaxy inside a radius $R$, the system is not completely
isolated and the effect of the matter at a radius $r>R$ has to be
considered as an unknown external pressure~\citep{chandrasekhar69}
(this term and the rotation velocity  are often neglected in the
case of elliptical galaxies). In order to overcome this problem,
it is convenient to compute the masses of galaxies starting from
the Jeans  equation that describes the equilibrium of a
spheroidally symmetric system having eccentricity $e$ and an
isotropic velocity dispersion tensor:
\begin{equation}
\frac{1}{\rho
(r)}\frac{d}{dr}[\rho(r)\sigma^2(r)]-\frac{v^2(r)}{r} = -
\frac{4\pi G(1-e^2)^{1/2}}{r}\int^r_o \frac{dx \
x^2\rho(x)}{(r^2-x^2e^2)^{1/2}}\ \ ,
\label{Eq_B1}
\end{equation}
where $r$ is the radius in the equatorial ($z=0$) plane, $\rho(r)$
is the (unknown) spatial density, $\sigma(r)$ and $v(r)$ are the
one--dimensional velocity dispersion and rotation velocity
respectively~\citep{binney87}. The solution of Equation~(\ref{Eq_B1})
can be written in the form $\rho(r)= \rho_0 \times l(r)$, where
$l(r)$ is the luminosity density. For elliptical galaxies the
approach followed for example by \citet{busarelloLongo92} is to
assume that the luminosity distribution corresponds to the spatial
deprojection of the $r^{1/4}$ law. A simple analytical
approximation for the deprojection of the $r^{1/4}$ law has been
derived by \citet{mellier87}:
\begin{equation}
l(r) = r^{-\beta} \exp{(-b \, r^{1/4})},
\label{Eq_B2}
\end{equation}
where $\beta =0.855$ and $b=7.669$. Substituting this solution
together with $V(r)$ and $\sigma(r)$ (deprojected from the
experimental data) in Equation~(\ref{Eq_B1}), an expression of
$\rho_0$ as a function of $r$ can be obtained. Computing the value
of $\rho_0$ for each object in the considered sample of galaxies
at 10 different radii, the residuals with respect to its mean
value $<\rho_0>$ turn out to be very small (and will be used to
estimate the error $\Delta M$), and show no systematic trend with
the radius, thus supporting the hypothesis that $\rho_0=constant$
at least in the inner regions~\citep{busarelloLongo92}. So, the
final result for the mass density is $\rho(r)= <\rho_0> r^{-\beta}
\exp(-b \, r^{1/4})$. Changing the luminosity distribution, this
method can be applied also to late type galaxies. A similar
approach was followed and well explained in their paper by
\citet{haring04}.

When the galaxy sample is very  large and various, it is very
difficult that all the masses of the galaxies are calculated in
the same manner. For the sample in  Table~\ref{Tab_01}, the masses
of most of the galaxies have been estimated with the procedure of
\citet{haring04}, a small part with the Schwarzschild
model~\citep{cappellari06}, and only in a few cases using the
dynamical mass formula.

\newpage
\begin{deluxetable}{lcccccccc}
\tabletypesize{\scriptsize} \tablecolumns{9} \tablecaption{Sample$^{\mathrm{a}}$} \tablewidth{0pt}
\tablehead{
\colhead{Galaxy}        &   \colhead{Type$^{\mathrm{b}}$} &   \colhead{$\sigma_{\mathrm{c}}$}     &   \colhead{References}        & \colhead{$M_{\bullet}$}    &   \colhead{$\delta M_{\bullet}$} & \colhead{References}                               &   \colhead{$M_\mathrm{G}$}    & \colhead{References}\\
\colhead{}              &   \colhead{}      & \colhead{(km/s)} & \colhead{}         &   \colhead{($M_{\sun}$)}     &
\colhead{($M_{\sun}$)}             & \colhead{}                            &   \colhead{($M_{\sun}$)}     & \colhead{}}
\startdata
CygnusA     &   E   &   270 & 1 &   $   2.5 \times 10^9 $   &   $   7.0 \times 10^8 $   & 1 & $   1.6 \times 10^{12}  $   & 2  \\
NGC221      &   E2  &   72  & 1 &   $   2.5 \times 10^6 $   &   $   5.0 \times 10^5 $   & 1 & $   8.0 \times 10^8     $   & 3  \\
NGC821      &   E6  &   200 & 1 &   $   8.5 \times 10^7 $   &   $   3.5 \times 10^7 $   & 1 & $   1.3 \times 10^{11}  $   & 3  \\
NGC1399     &   E1  &   329 & 1 &   $   4.8 \times 10^8 $   &   $   7.0 \times 10^7 $   & 1 & $   2.32 \times 10^{11} $   & 4  \\
NGC2974     &   E4  &   227 & 1 &   $   1.7 \times 10^8 $   &   $   3.0 \times 10^7 $   & 1 & $   1.57 \times 10^{11} $   & 5   \\
NGC3377     &   E5  &   139 & 1 &   $   8.0 \times 10^7 $   &   $   6.0 \times 10^6 $   & 1 & $   3.08 \times 10^{10} $   & 5   \\
NGC3379     &   E1  &   207 & 1 &   $   1.4 \times 10^8 $   &   $   2.7 \times 10^8 $   & 1 & $   6.8 \times 10^{10}  $   & 3  \\
NGC3608     &   E2  &   192 & 1 &   $   1.9 \times 10^8 $   &   $   1.0 \times 10^8 $   & 1 & $   9.7 \times 10^{10}  $   & 3  \\
NGC4261     &   E2  &   309 & 1 &   $   5.2 \times 10^8 $   &   $   1.1 \times 10^8 $   & 1 & $   3.6 \times 10^{11}  $   & 3  \\
NGC4291     &   E2  &   285 & 1 &   $   3.1 \times 10^8 $   &   $   2.3 \times 10^8 $   & 1 & $   1.3 \times 10^{11}  $   & 3  \\
NGC4374     &   E1  &   281 & 1 &   $   4.64 \times 10^8$   &   $   3.46 \times 10^8$   & 1 & $   3.6 \times 10^{11}  $   & 3  \\
NGC4473     &   E5  &   179 & 1 &   $   1.1 \times 10^8 $   &   $   8.0 \times 10^7 $   & 1 & $   9.2 \times 10^{10}  $   & 3  \\
NGC4486     &   E0  &   332 & 1 &   $   3.4 \times 10^9 $   &   $   1.0 \times 10^9 $   & 1 & $   6.0 \times 10^{11}  $   & 3  \\
NGC4486A    &   E2  &   110 & 1 &   $   1.3 \times 10^7 $   &   $   8.0 \times 10^6 $   & 1 & $   4.06 \times 10^9     $   & 6  \\
NGC4621     &   E5  &   225 & 1 &   $   4.0 \times 10^8 $   &   $   6.0 \times 10^7 $   & 1 & $   1.88 \times 10^{11} $   & 5 \\
NGC4649     &   E1  &   335 & 1 &   $   2.0 \times 10^9 $   &   $   6.0 \times 10^8 $   & 1 & $   4.9 \times 10^{11}  $   & 3  \\
NGC4697     &   E4  &   174 & 1 &   $   1.7 \times 10^8 $   &   $   2.0 \times 10^7 $   & 1 & $   1.1 \times 10^{11}  $   & 3  \\
NGC5077     &   E3  &   255 & 1 &   $   7.4 \times 10^8 $   &   $   4.7 \times 10^8 $   & 1 & $   2.1 \times 10^{11}  $   & 7  \\
NGC5813     &   E1  &   239 & 1 &   $   7.0 \times 10^8 $   &   $   1.1 \times 10^8 $   & 1 & $   5.05 \times 10^{11} $   & 5  \\
NGC5845     &   E3  &   233 & 1 &   $   2.4 \times 10^8 $   &   $   1.4 \times 10^8 $   & 1 & $   3.7 \times 10^{10}  $   & 3  \\
NGC5846     &   E0  &   237 & 1 &   $   1.1 \times 10^9 $   &   $   2.0 \times 10^8 $   & 1 & $   6.36 \times 10^{11} $   & 5  \\
NGC6251     &   E2  &   311 & 1 &   $   5.9 \times 10^8 $   &   $   2.0 \times 10^8 $   & 1 & $   5.6 \times 10^{11}  $   & 3  \\
NGC7052     &   E4  &   277 & 1 &   $   3.7 \times 10^8 $   &   $   2.6 \times 10^8 $   & 1 & $   2.9 \times 10^{11}  $   & 3  \\
NGC3115     &   S0  &   252 & 1 &   $   9.1 \times 10^8 $   &   $   1.03 \times 10^9$   & 1 & $   1.2 \times 10^{11}  $   & 3  \\
NGC3245     &   S0  &   210 & 1 &   $   2.1 \times 10^8 $   &   $   5.0 \times 10^7 $   & 1 & $   6.8 \times 10^{10}  $   & 3  \\
NGC3414     &   S0  &   237 & 1 &   $   2.5 \times 10^8 $   &   $   4.0 \times 10^7 $   & 1 & $   1.7 \times 10^{11}  $   & 5  \\
NGC3998     &   S0  &   305 & 1 &   $   2.2 \times 10^8 $   &   $   2.0 \times 10^8 $   & 1 & $   5.5 \times 10^{10}  $   & 8  \\
NGC4342     &   S0  &   253 & 1 &   $   3.3 \times 10^8 $   &   $   1.9 \times 10^8 $   & 1 & $   1.2 \times 10^{10}  $   & 3  \\
NGC4459     &   S0  &   178 & 1 &   $   7.0 \times 10^7 $   &   $   1.3 \times 10^7 $   & 1 & $   7.86 \times 10^{10} $   & 5 \\
NGC4552     &   S0  &   252 & 1 &   $   4.8 \times 10^8 $   &   $   8.0 \times 10^7 $   & 1 & $   1.87 \times 10^{11} $   & 5  \\
NGC4564     &   S0  &   157 & 1 &   $   5.6 \times 10^7 $   &   $   3.0 \times 10^6 $   & 1 & $   4.4 \times 10^{10}  $   & 3  \\
NGC5128     &   S0  &   120 & 1 &   $   4.9 \times 10^7 $   &   $   1.8 \times 10^7 $   & 1 & $   2.16 \times 10^{10} $   & 9  \\
NGC5252     &   S0  &   190 & 1 &   $   1.06\times 10^9 $   &   $   1.63 \times 10^9$   & 1 & $   2.4 \times 10^{11}  $   & 2  \\
NGC1023     &   SB0 &   204 & 1 &   $   4.4 \times 10^7 $   &   $   5.0 \times 10^6 $   & 1 & $   6.9 \times 10^{10}  $   & 3  \\
NGC2778     &   SB0 &   162 & 1 &   $   1.4 \times 10^7 $   &   $   9.0 \times 10^6 $   & 1 & $   1.06 \times 10^{10} $   & 10  \\
NGC2787     &   SB0 &   210 & 1 &   $   4.1 \times 10^7 $   &   $   5.0 \times 10^6 $   & 1 & $   2.9 \times 10^{10}  $   & 11  \\
NGC3384     &   SB0 &   148 & 1 &   $   1.6 \times 10^7 $   &   $   2.0 \times 10^6 $   & 1 & $   2.0 \times 10^{10}  $   & 3  \\
NGC4596     &   SB0 &   149 & 1 &   $   7.9 \times 10^7 $   &   $   3.8 \times 10^7 $   & 1 & $   2.6 \times 10^{10}  $   & 2  \\
Circinus    &   S   &    75 & 1 &   $   1.1 \times 10^6 $   &   $   2.0 \times 10^5 $   & 1 & $   3.0 \times 10^9     $   & 12  \\
NGC224      &   SA  &   170 & 1 &   $   1.4 \times 10^8 $   &   $   9.0 \times 10^7 $   & 1 & $   4.4 \times 10^{10}  $   & 13  \\
NGC3031     &   SA  &   162 & 1 &   $   7.6 \times 10^7 $   &   $   2.2 \times 10^7 $   & 1 & $   1.0 \times 10^{10}  $   & 14  \\
MW          &   SB  &   100 & 1 &   $   3.7 \times 10^6 $   &   $   2.0 \times 10^5 $   & 1 & $   1.1 \times 10^{10}  $   & 3  \\
NGC1300     &   SB  &   229 & 1 &   $   7.3 \times 10^7 $   &   $   6.9 \times 10^7 $   & 1 & $   2.14 \times 10^{10} $   & 15  \\
NGC3079     &   SB  &   146 & 1 &   $   2.4 \times 10^6 $   &   $   2.4 \times 10^6 $   & 1 & $   1.7 \times 10^9     $   & 16  \\
NGC3227     &   SAB &   133 & 1 &   $   1.4 \times 10^7 $   &   $   1.0 \times 10^7 $   & 1 & $   2.95 \times 10^9    $   & 17  \\
NGC4151     &   SAB &   156 & 1 &   $   6.5 \times 10^7 $   &   $   7.0 \times 10^6 $   & 1 & $   1.09 \times 10^{11} $   & 18 \\
NGC4258     &   SAB &   134 & 1 &   $   3.9 \times 10^7 $   &   $   1.0 \times 10^6 $   & 1 & $   1.1 \times 10^{10}  $   & 2  \\
NGC4945     &   SB  &   100 & 1 &   $   1.4 \times 10^6 $   &   $   1.4 \times 10^6 $   & 1 & $   3.0 \times 10^9     $   & 12  \\
NGC7582     &   SB  &   156 & 1 &   $   5.5 \times 10^7 $   &   $   2.6 \times 10^7 $   & 1 & $   1.31 \times 10^{11} $   & 19  \\
\hline
IC1459      &   E3  &   306 & 1 &   $   1.5 \times 10^9 $   &   $   1.0 \times 10^9 $   & 2 &   $   6.6  \times 10^{11} $   & 2 \\
IC4296      &   E   &   336 & 1 &   $   1.3 \times 10^9 $   &   $   4.0 \times 10^8 $   & 1 &   $   1.56 \times 10^{12} $   & 20    \\
NGC3607$^{\mathrm{c}}$ & E   &   229 & 21 &  $   1.2 \times 10^8 $   &   $   4.0 \times 10^7 $   & 21 &  $    2.70 \times 10^{11}$   & 21 \\
NGC4486B    &   E0  &   169 & 1 &   $   6.0 \times 10^8 $   &   $   3.0 \times 10^8 $   & 1 &   $   1.22 \times 10^{11} $   & 22    \\
NGC4742     &   E4  &   109 & 1 &   $   1.4 \times 10^7 $   &   $   5.0 \times 10^6 $   & 1 &   $   6.2 \times 10^9  $      & 3 \\
NGC5576$^{\mathrm{c}}$ & E3  &   183 & 21 &  $   1.8 \times 10^8 $   &   $   4.0 \times 10^7 $   & 21 &  $    1.47 \times 10^{11}$   & 21 \\
NGC3585$^{\mathrm{c}}$ & S0  &   213 & 21 &  $   3.4 \times 10^8 $   &   $   1.5 \times 10^8 $   & 21 &  $    1.85 \times 10^{11}$   & 21 \\
NGC4026$^{\mathrm{c}}$ & S0  &   180 & 21 &  $   2.1 \times 10^8 $   &   $   7.0 \times 10^7 $   & 21 &  $    5.17 \times 10^{10}$   & 21 \\
NGC7332     &   S0  &   135 & 1 &   $   1.3 \times 10^7 $   &   $   6.0 \times 10^6 $   & 1 &   $   1.5 \times 10^{10}  $   & 3 \\
NGC7457     &   S0  &   69  & 1 &   $   3.5 \times 10^6 $   &   $   1.4 \times 10^6 $   & 1 &   $   7.0 \times 10^9  $      & 3 \\
NGC4203     &   SB0 &   124 & 11 &  $   5.2 \times 10^7 $   &   $   1.0 \times 10^6 $   & 11 &  $   1.5 \times 10^{10}  $   & 11 \\
NGC1068     &   SA  &   151 & 1 &   $   8.4 \times 10^6 $   &   $   3.0 \times 10^5 $   & 1 &   $   1.5 \times 10^{10}  $   & 23    \\
NGC2748     &   SA  &   79  & 24 &   $   4.4 \times 10^7 $   &   $   3.6 \times 10^7 $   & 15 &   $   1.69 \times 10^{10} $   & 15 \\
NGC4594     &   SA  &   240 & 1 &   $   1.0 \times 10^9 $   &   $   1.0 \times 10^9 $   & 3 &   $    2.7 \times 10^{11} $   & 3 \\
NGC7469     &   SAB &   152 & 25 &   $   1.22 \times 10^7$   &   $   1.40 \times 10^6    $   & 25 &   $   4.5 \times 10^{9}   $   & 26
\enddata
\tablerefs{
(1) \citet{graham08};
(2) \citet{marconi03};
(3) \citet{haring04};
(4) \citet{houghton06};
(5) \citet{cappellari06}, \citet{cappellari09};
(6) \citet{nowak07};
(7) \citet{defrancesco08};
(8) \citet{defrancesco06};
(9) \citet{bekki03};
(10) \citet{aller07};
(11) \citet{sarzi01};
(12) \citet{hitschfeld08};
(13) \citet{riffeser08};
(14) \citet{sofue98};
(15) \citet{atkinson05};
(16) \citet{koda02};
(17) \citet{wandel02};
(18) \citet{wandel99};
(19) \citet{wold06};
(20) \citet{dallabonta07};
(21) \citet{gultekin09a};
(22) \citet{bacon85};
(23) \citet{israel09};
(24) \citet{batcheldor05};
(25) \citet{hicks08};
(26) \citet{genzel95}.
}
\tablecomments{$^{\mathrm{a}}$Adopting the same strategy as in Paper II, we consider that the error for the bulge mass is 0.18 dex in $\log_{10}M_{\mathrm{G}}$ for all the galaxies, while the relative error on the velocity dispersions is $10\%$. $^{\mathrm{b}}$Galaxy types are taken from \citet{graham08} with the following exceptions: IC4296, NGC221, NGC821, NGC1399, NGC2974, NGC3379, NGC3607, NGC4374, NGC4486A, NGC4486B, NGC4621, NGC5077, NGC5813, NGC5846, NGC6251, which are taken from the NASA/IPAC Extragalactic database. $^{\mathrm{c}}$Effective dispersion $\sigma_{\mathrm{e}}$ is given rather than the central $\sigma_{\mathrm{c}}$.}
\label{Tab_01}
\end{deluxetable}

\newpage
\begin{deluxetable}{cccccccc}
\tabletypesize{\scriptsize} \tablecolumns{8} \tablecaption{Fitting parameters of the $M_{\bullet} - M_{\mathrm{G}} \sigma^2$, $M_{\bullet} -\sigma$ and $M_{\bullet} - M_{\mathrm{G}}$ relations, both for sample A and B}
\tablewidth{0pt}
\tablehead{
\colhead{Relation} & \colhead{Sample} & \colhead{$N$} & \colhead{$m\pm\Delta m$} &
\colhead{$b\pm\Delta b$} & \colhead{$\chi_{\mathrm{r}}^2$} & \colhead{$\varepsilon_{0}$} & \colhead{$r$} \\
(1) & (2)& (3)& (4)& (5) & (6) & (7)& (8)}
\startdata
$M_{\bullet}-M_{\mathrm{G}}\sigma^2$ & A  &   49  &   $0.80\pm0.03$ &   $4.49\pm0.15$  &   1.74   &0.19&   0.92    \\
$M_{\bullet}-\sigma$                 & A  &   49  &   $5.06\pm0.25$ &   $8.18\pm0.04$  &   1.85   &0.25&   0.87    \\
$M_{\bullet}-M_{\mathrm{G}}$         & A  &   49  &   $1.15\pm0.05$ &   $-4.35\pm0.57$ &   2.10   &0.27&   0.89    \\
$M_{\bullet}-M_{\mathrm{G}}\sigma^2$ & B  &   64  &   $0.80\pm0.03$ &   $4.53\pm0.13$  &   1.92   &0.21&   0.92    \\
$M_{\bullet}-\sigma$                 & B  &   64  &   $5.00\pm0.21$ &   $8.20\pm0.04$  &   2.34   &0.32&   0.85    \\
$M_{\bullet}-M_{\mathrm{G}}$         & B  &   64  &   $1.13\pm0.05$ &   $-4.19\pm0.49$ &   2.77   &0.26&   0.90    \\
\enddata
\tablecomments{We report the used relation in Column 1, the sample in Column 2, and
the corresponding number of galaxies in Column 3. By using the routine FITEXY, we find the best
fit of the relationship $y = b + m x$. The results for $m$ and $b$ are in
Columns 4 and 5 and the corresponding reduced $\chi_{\mathrm{r}}^2=\chi^2/(N-2)$ in Column 6. The internal (intrinsic)
scatter (Column 7) is estimated as that which yields a $\chi_{\mathrm{r}}^2=1$ with respect to the given best-fitting relation (see also Appendix~\ref{appendix}). Finally, the linear correlation coefficient is shown in Column 8.}
\label{Tab_02}
\end{deluxetable}
\newpage
%
\begin{deluxetable}{cccccc}
\tabletypesize{\scriptsize} \tablecolumns{6} \tablecaption{Fitting parameters of the $M_{\bullet}-M_{\mathrm{G}} \sigma^2$, $M_{\bullet} -\sigma$ and $M_{\bullet} - M_{\mathrm{G}}$ relations, both for sample A and B, without considering errors in the kinetic energy}
\tablewidth{0pt}
\tablehead{
\colhead{Relation} & \colhead{Sample} & \colhead{$N$} & \colhead{$m\pm\Delta m$} &
\colhead{$b\pm\Delta b$} & \colhead{$\varepsilon_{\mathrm{y}}^2$} \\
(1) & (2)& (3)& (4)& (5) & (6)}
\startdata
$M_{\bullet}-M_{\mathrm{G}}\sigma^2$ & A  &   49  &   $0.74\pm0.04$ &   $4.80\pm0.20$ &   0.10   \\
$M_{\bullet}-\sigma$ & A  &   49  &   $4.46\pm0.36$ &   $8.13\pm0.06$ &   0.16   \\
$M_{\bullet}-M_{\mathrm{G}}$ & A  &   49  &   $0.98\pm0.07$ &   $-2.56\pm0.79$ &   0.15 \\
$M_{\bullet}-M_{\mathrm{G}}\sigma^2$ & B  &   64  &   $0.73\pm0.04$ &   $4.88\pm0.18$ &   0.11   \\
$M_{\bullet}-\sigma$ & B  &   64  &   $4.12\pm0.32$ &   $8.17\pm0.06$ &   0.19 \\
$M_{\bullet}-M_{\mathrm{G}}$ & B  &   64  &   $0.99\pm0.06$ &   $-2.60\pm0.67$ &   0.14   \\
\enddata
\tablecomments{We report the used relation in Column 1, the sample in Column 2, and
the corresponding number of galaxies in Column 3. By using a standard least squared fitting and assuming that errors in
the galaxy mass and velocity dispersion are zero and that errors in the $\log_{10} M_{\bullet}$ are the same $\epsilon_{\mathrm{y}}$ for
each galaxy, we find the best fit of the relationship $y = b + m x$. The results for $m$ and $b$ are in Columns 4 and 5, respectively. The values of the $\epsilon_{\mathrm{y}}$ are in Column 6.}
\label{Tab_03}
\end{deluxetable}

\newpage
\begin{deluxetable}{cccccc}
\tabletypesize{\scriptsize} \tablecolumns{6} \tablecaption{Fitting parameters of the $M_{\bullet} - M_{\mathrm{G}} \sigma^2$, $M_{\bullet} -\sigma$ and $M_{\bullet} - M_{\mathrm{G}}$ relations, for sample A, with a different method to calculate the internal scatter}
\tablewidth{0pt}
\tablehead{
\colhead{Relation} & \colhead{Sample} & \colhead{$N$}  & \colhead{$\varepsilon_{0}$} & \colhead{$m\pm\Delta m$} &
\colhead{$b\pm\Delta b$} \\
(1) & (2)& (3)& (4)& (5) & (6)}
\startdata
$M_{\bullet}-M_{\mathrm{G}}\sigma^2$ & A  &   49  &   0.110  &  $0.78\pm0.04$ &   $4.61\pm0.19$    \\
$M_{\bullet}-\sigma$                 & A  &   49  &   0.230  &  $4.25\pm0.35$ &   $8.18\pm0.06$    \\
$M_{\bullet}-M_{\mathrm{G}}$         & A  &   49  &   0.165  &  $1.07\pm0.07$ &   $-3.55\pm0.78$   \\
\enddata
\tablecomments{We report the used relation in Column 1, the
sample in Column 2, and the corresponding number  of galaxies in
Column 3. The values of the internal scatter in Column 4 are
calculated with the procedure explained in
Appendix~\ref{appendix}. The corresponding values for the slope
and the normalization are reported in Columns 5 and 6,
respectively.} \label{Tab_04}
\end{deluxetable}

\newpage

\newpage

\newpage

\end{document}